\documentclass[10pt,letterpaper,twocolumn]{article} 

\usepackage{ol2}
\usepackage[draft,implicit=false]{hyperref}
\usepackage{amsmath}
\usepackage{amssymb}
\usepackage{color}
\usepackage[english]{babel}

\begin{document}

\twocolumn[ 

\title{An all-optical buffer based on \\temporal cavity solitons operating at 10~Gb/s}

\author{Jae K. Jang,$^1$ Miro Erkintalo,$^1$ Jochen Schr\"oder,$^2$ \\Benjamin J. Eggleton,$^3$ Stuart G. Murdoch,$^{1,*}$ and St\'ephane Coen$^1$}
\address{
$^1$Dodd-Walls Centre for Photonic and Quantum Technologies, and Physics Department, The University of Auckland, Private~Bag~92019, Auckland 1142, New Zealand \\
$^2$School of Electronic and Computer Systems Engineering, RMIT University, Melbourne, VIC3000, Australia \\
$^3$Centre for Ultrahigh bandwidth Devices for Optical Systems (CUDOS), Institute of Photonics and Optical Science (IPOS), School of Physics, University of Sydney, NSW 2006, Australia \\
$^*$Corresponding author: s.murdoch@auckland.ac.nz
}

\begin{abstract}
We demonstrate the operation of an all-optical buffer based on temporal cavity solitons stored in a nonlinear passive fiber ring resonator. Unwanted acoustic interactions between neighboring solitons are suppressed by modulating the phase of the external laser driving the cavity. A new locking scheme is presented that allows the buffer to operate with an arbitrarily large number of cavity solitons in the loop. Experimentally, we are able to demonstrate the storage of 4536~bits of data, written all-optically into the fiber ring at 10~Gb/s, for 1~minute.
\end{abstract}


 ]

\noindent The vision of what will constitute a future optical network has evolved considerably in recent years~\cite{Saleh:06,Saleh:12, Tomkos:14, Napoli:15}. Nonetheless, the ability to write optical data directly into an optical storage buffer remains a highly desirable functional component in a truly all-optical network. Optical buffers based on slow light materials~\cite{Chang-Hasnain:03,Tucker:05}, optical delay lines~\cite{Langenhorst:96,Tanemura:11}, parametric effects~\cite{Agrawal:05}, coupled-resonators~\cite{Xia:07,Melloni:08} and optical solitons~\cite{McDonald:90,Firth:96,Barland:02,Leo:10,Garbin:15} have all been proposed and demonstrated. Each of these implementations comes with its own distinct set of advantages and challenges. In this Letter we concentrate on optical buffers based on cavity solitons (CS)~\cite{McDonald:90}. CSs are dissipative solitons found in nonlinear optical cavities driven by an external coherent field. They have attracted considerable interest thanks to their very long storage times~\cite{Jang:13,Jang:15}, scalable geometries~\cite{Herr:14}, and their potential to be selectively, and independently, written or erased~\cite{Barland:02,Jang_writing:15}. They are also robust attractors making them easy to excite and very stable against external perturbations~\cite{Leo:10}. In the spatial domain, CSs take the form of spatially confined peaks in the intensity of the intra-cavity field, with their spatial profile set to ensure an exact balance between the gain, dissipation, diffraction, and nonlinearity of the cavity~\cite{McDonald:90,Firth:96,Barland:02}. In the temporal domain, dispersion takes the place of diffraction, and CSs manifest themselves as pulses of light that circulate in an optical resonator~\cite{Leo:10,Jang:13}. Optical memories based on spatial CSs were first proposed in the 1990s~\cite{McDonald:90}. However, to date, their storage capacity has been limited to only a few bits~\cite{Barland:02}. In 2010 Leo et. al, reported the first demonstration of an optical buffer based on temporal CSs~\cite{Leo:10}. Using a 380~m loop of standard optical single-mode fiber~(SMF) they were able to demonstrate a few seconds storage of $\sim 15$~optical bits written into the cavity as a sequence of temporal CSs at a bit-rate of 10~Mb/s. More recently, a CS buffer storing 21 optical bits at a bit rate of 10~Gb/s with a storage time of a few minutes has been reported~\cite{Jang:15}. To date, however, studies of temporal CSs have focussed on unveiling their characteristics and dynamics, whilst their practical buffering capabilities have not been thoroughly explored.

In this Letter, we consider the performance limits of a temporal CS-based buffer in terms of storage capacity, bit-rate and storage time. We also introduce a modified locking circuit capable of maintaining the driving field at the required detuning even when the intra-cavity power of the loop increases significantly due to the large number of CS present in a full buffer. This considerably improves on the performance of the buffers originally reported in Refs.~\cite{Leo:10,Jang:15} with 4536~bits of optical data written directly into the cavity at a rate of 10~Gb/s. The data is then stored, in the form of temporal CSs, for 1~minute before being recovered with a fidelity of 100\%. The temporal CS buffer we present here belongs to the class of macroscopic optical buffers capable of storing a large number of bits for long periods of time. Recently, important connections have been highlighted between the nonlinear physics of fiber rings and optical microresonators, with CSs observed in both geometries~\cite{Leo:10,Herr:14}. This raises the prospect of CS buffers utilizing micron-scale resonators, which would allow for the significant miniaturization of these devices, potentially down to the level of the coupled-resonator-optical-waveguide (CROW) buffers reported in Ref.~\cite{Xia:07}.

\begin{figure}[t]
\centerline{\includegraphics[width=0.9\columnwidth]{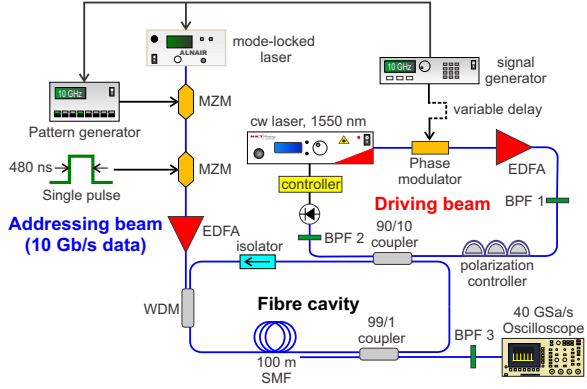}}
\caption{A schematic of the temporal CS-based all-optical buffer. EDFA: erbium-doped fiber amplifier, BPF: band-pass filter, SMF: single-mode fiber, WDM: wavelength-division-multiplexer, MZM: Mach-Zehnder modulator.}
\label{fig:1}
\end{figure}

The optical fiber ring we use for our all-optical buffer, is based on designs previously used to study the dynamics of temporal CSs ~\cite{Leo:10,Jang:13}. A schematic diagram is shown in Fig.~\ref{fig:1}. The cavity consists of a 90/10 fiber coupler and 100~m of standard SMF (roundtrip time $t_{\mathrm{R}} \simeq 480$~ns). It also incorporates a fiber isolator to suppress the growth of any counter-propagating field induced by stimulated Brillouin scattering, and a wavelength-division multiplexer (WDM) through which optical addressing pulses can enter the cavity. Moreover the intra-cavity dynamics can be directly monitored through the 1\% output port of an additional 99/1 fiber coupler. Overall the cavity has a measured finesse of 22 and exhibits $\sim 100$~kHz resonance width. The driving beam is provided by a narrow linewidth ($< 1$~kHz) distributed feedback fiber laser at 1550~nm which is amplified to $\sim 1$~W average power with an Erbium-doped fiber amplifier (EDFA). The amplifier noise is removed with a 75~GHz full-width-at-half-maximum (FWHM) optical band-pass filter (BPF) also centered at 1550~nm (labelled BPF~1 in Fig.~\ref{fig:1}) before the filtered driving beam enters the cavity through the 90/10 coupler.

\begin{figure}[t]
\centerline{\includegraphics[width=0.7\columnwidth,clip=true]{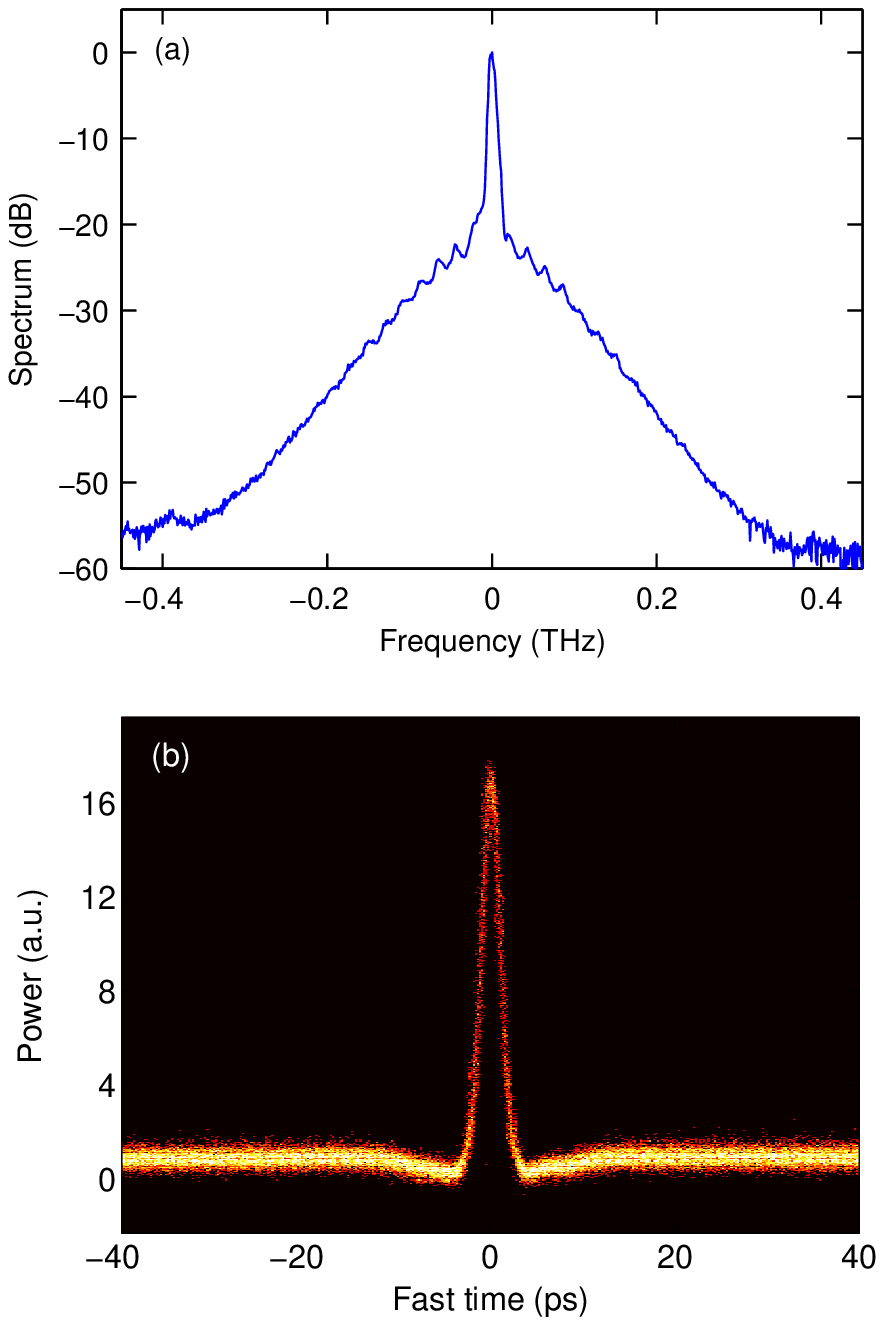}}
\caption{(a) The spectral profile and (b) the temporal profile of an isolated CS.}
\label{fig:2}
\end{figure}

To allow for stable operation of the buffer, the frequency of the driving beam needs to be locked at a constant phase detuning from a mode of the cavity by a feedback control loop~\cite{Leo:10,Jang:13}. In previous studies researchers have used the average power reflected off the cavity to generate an error signal for their loop. Locking this signal to a constant value locks the detuning. However, these schemes do not work when a significant number of CSs are present. This is because a CS is a pulse of high intensity light that rides on top of the CW background. Thus, writing a large number of CSs into the cavity actually increases the average power of the intra-cavity beam, disrupting the operation of the feedback circuit. More quantitatively, in our system a temporal CS manifests itself as a 2.6~ps pulse with a peak power of $\sim 10$~W embedded on a $\sim 0.87$~W continuous-wave~(cw) background radiation. For a cavity filled with CSs at 10 GHz (480 ns x 10 GHz = 4800 CS) the total energy carried by the CSs amounts to $\sim$ 125~nJ which is comparable with the energy contained in the background field over one cavity roundtrip ($\sim 418$~nJ). Writing this number of CSs into the ring results in a severe perturbation to a locking signal derived from the average power. To circumvent this problem, we add a programmable optical filter (WaveShaper~1000E) just before the photodiode that measures the power reflected off the ring. The filter is configured to act as a 20~GHz wide BPF (hence labelled BPF~2 in Fig.~\ref{fig:1}) centered about the driving wavelength. This means the photodiode measures only the power of the cw component of the intra-cavity beam with negligible contribution from the spectrally broad ($\sim$ 1nm FWHM) CS. This modification allows the locking circuit to operate correctly with an arbitrarily large number of CS in the cavity. The locking will also remain unaffected if the number of CSs change as bits are written into, or erased from, the cavity~\cite{Jang_writing:15}. In the experiments that follow the cavity detuning is locked to $\delta_0 \simeq 0.41$~rad using this method. CSs are written onto the intra-cavity beam via cross-phase-modulation from an external addressing beam consisting of optical pulses (1.8~ps FWHM, $\sim 10$~W peak power) generated by a 10~GHz mode-locked laser centered at 1532~nm. We find this method of writing CSs to have a higher fidelity than the alternative phase modulation scheme presented in Ref.~\cite{Jang_writing:15}. More details of this writing scheme are provided in Ref.~\cite{Leo:10}. The addressing beam is coupled into the cavity via the intra-cavity WDM. It circulates only once around the loop before exiting via the same WDM. A single addressing pulse is sufficient to write a CS into the loop. Once written, the CS will circulate as long as the driving beam is maintained~\cite{Lugiato:03}. In Figs.~\ref{fig:2}(a) and (b) we show the measured spectral and temporal profiles of an isolated CS. The spectral measurement is made using a 0.05~nm resolution optical spectrum analyzer, and the temporal measurement with a picosecond resolution optical sampling scope (EXFO PSO100). The spectral profile shows the CS spectrum ($\sim 0.14$~THz FWHM) centered around the cw peak of the driving beam located at 1550~nm. The temporal profile exhibits the characteristic form of a CS: a short pulse (2.6~ps FWHM), with small dips on either side, riding on the cw background of the intra-cavity beam~\cite{Leo:10}.

In Ref.~\cite{Jang:13} it was reported for the first time that neighboring temporal CSs can interact via acoustic waves generated by electrostriction~\cite{Dianov:90}. These acoustic waves result in a leading CS inducing a minute change in the velocities of any trailing CSs over a range of separations up to 100~ns. Measurements and calculations show that the magnitude of this interaction is extremely weak with the trailing solitons changing their temporal positions in the buffer by at most several tens of attoseconds per microsecond of storage time~\cite{Jang:13}. Nonetheless, for a high bit-rate optical buffer with storage times of the order of seconds or longer this effect must be fully suppressed to prevent the data bits from walking out of their assigned bit slots. We achieve this suppression by providing a sinusoidal phase modulation to the input driving field~\cite{Firth:96} using a 10~GHz optical modulator~\cite{Jang:15}. This imprints onto the intra-cavity beam a sinusoidal phase profile whose local maxima act as robust attractive trapping sites for a CS. As a consequence, the CS drifts towards the phase maximum at a rate of $\delta\tau(\tau) = -\beta_2 L\phi'(\tau)$ per roundtrip, where $\beta_2$ is the cavity group velocity dispersion, $L$ is the cavity length, and $\phi'(\tau)$ is the gradient of the phase modulation at the position of the CS~\cite{Jang:15}. For the cavity presented in this work $L = 100$~m, $\beta_2 = -21.4~\mathrm{ps^2/km}$, and the amplitude and frequency of the phase modulation are $A \simeq 0.3$~rad and $f_{\mathrm{PM}}\simeq 10$~GHz respectively. This gives a maximum phase-modulation-induced drift rate of $\sim 20$~fs/roundtrip ($\sim 40$~fs/$\mu$s), more than 3 orders of magnitude higher than that of the acoustic interaction, ensuring complete suppression of this unwanted acoustic effect. An additional advantage of the technique is that the trapped CS will be automatically re-timed to the frequency of the phase modulation $f_{\mathrm{PM}}$ which in this experiment we set to the clock frequency of the input data. We note that this level of phase modulation is not large enough to directly write CSs into the cavity~\cite{Jang_writing:15}.

For correct operation of the buffer the frequency of the sinusoidal phase modulation must be set to match both the bit rate of the input data, and an integer multiple of the cavity's free-spectral-range ($f_{\mathrm{PM}} = m\cdot FSR$). This second requirement ensures the phase modulation of the driving beam is efficiently coupled into the cavity. In our study we set $f_{\mathrm{PM}} \simeq 9.83$~GHz. This frequency corresponds to 4736~times the FSR of the fiber loop (i.e. $m = 4736$) and sets the number of available bit periods in the buffer. As shown in Ref.~\cite{Jang:15}, this frequency does not have to be actively locked to the FSR of the cavity as the phase trapping mechanism still works with a small mismatch between the modulation frequency and $m\cdot FSR$. For the parameters of our cavity, the allowed mismatch is $\sim \pm 800$~Hz. To write an optical data sequence into the cavity, an NRZ electrical signal from a 10~Gb/s pattern generator is used to imprint data onto the optical output of the 10~GHz mode-locked laser via a Mach-Zehnder modulator (MZM). A single cycle of this pattern is then selected by a second MZM and coupled into the cavity. Each mode-locked pulse in this optical data stream writes its own CS into the cavity, thus encoding the data as a train of CSs. The input data signal is temporally synchronised with the phase modulated driving beam by means of a variable delay line (see Fig.~\ref{fig:1}, before the phase modulator) such that the position at which the CSs are written and the phase modulation peaks coincide. This ensures each CS is centered on its own attractive phase potential, and is not affected by acoustic induced interactions with adjacent CSs, or by any other weak perturbations. This also guarantees that the solitons are continuously re-timed to the $9.83$~GHz clock that drives the phase modulator. We demonstrate the operation of the buffer with a 4536~bit data sequence that fills 96\% of the available bit periods in the cavity. The remaining 200~bits periods are left empty. This is not technically necessary but helps distinguish one roundtrip of data from the next in the figures that follow.

\begin{figure}[t!]
\includegraphics[width=0.8\columnwidth,clip=true]{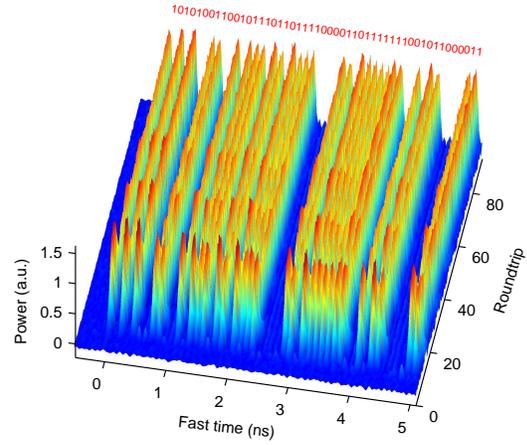}
\caption{First 5~ns of data stored as CSs measured at the 1\% tap coupler immediately after writing.}
\label{fig:3}
\end{figure}

\begin{figure}[b!]
\includegraphics[width=0.8\columnwidth,clip=true]{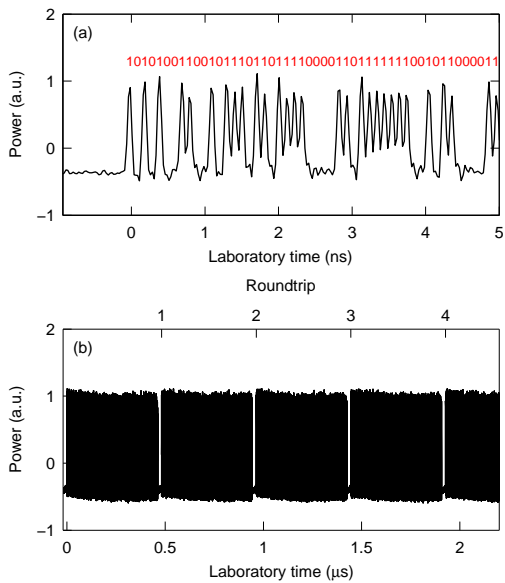}
\caption{(a) First 5~ns of data stored as CSs measured at the 1\% tap coupler 1~minute after writing. (b) CS encoded data (4536 bits out of the available 4736 used) measured over 4 roundtrips.}
\label{fig:4}
\end{figure}

Firstly, we investigate the formation dynamics of the CSs in the buffer. Figure~\ref{fig:3} shows the intra-cavity intensity, measured at the 1\% intra-cavity tap coupler, immediately after the application of the addressing pulses. A BPF (BPF~3 in Fig.~\ref{fig:1}, 1551~nm center wavelength, 75~GHz FWHM) removes the cw component of the output field before detection, resulting in an improved signal to noise ratio for the detected photocurrent~\cite{Jang:13}. For clarity, only the first 5~ns (50~bits) of the buffer is shown. Each addressing pulse is seen to write its own CS, with $\sim 30$~roundtrips required for the soliton to fully stabilize. This signal was measured using a 12~GHz photodiode and a real time oscilloscope. As a result, the detected pulses have the $\sim 50$~ps impulse response of this system rather than the true $2.6$~ps pulse width of the CSs. In Fig.~\ref{fig:4}(a) we show the same 5~ns of data measured 1~minute after the data was written into the loop. The recovered data signal is extremely clear with very good contrast between a logical `1' (a CS) and a logical `0' (no CS). In Fig.~\ref{fig:4}(b) we plot the entire optical data sequence, again measured 1~minute after it was initially written into the cavity, over 4~roundtrips ($4 \times 480~\mathrm{ns}$). The 200~unfilled bit periods at the end of each roundtrip are clearly visible. An investigation of the error-rate of this buffer shows that, after careful optimisation, in 8 out of 10 trials, the fidelity of the stored CS data sequence remains at 100\% (that is all 4536~bits are correctly recovered after 1~minute of storage). In the remaining two trials we typically find 1 or 2~errors in the stored data after 1~minute storage, with these errors predominantly occurring during the writing phase rather than the storage phase. This result represents an impressive demonstration of all-optical storage, with 1~minute of storage time corresponding to more that $10^8$~roundtrips of the fiber loop, or equivalently 12~million kilometers propagation distance.

Finally, we consider the performance limits of optical buffers based on CSs in macroscopic fiber rings. The existence of weak acoustic interactions between even widely spaced CSs necessitates phase modulation of the driving beam~\cite{Jang:15}. This requires the bit-rate of the buffer to match the phase modulation frequency, but sets no fundamental upper limit. In fact for a given phase modulation amplitude, the strength of the phase modulation trapping increases linearly with the modulation frequency~\cite{Jang:15}. The ultimate limit for the bit-rate of such a buffer is set by optical soliton-to-soliton interactions~\cite{Gordon:83} that occur when neighboring CSs approach to within a few pulse widths of each other. Here the strength of these optical interactions can overwhelm the re-timing capability of the phase modulated driving beam and adjacent CSs can merge or annihilate each other~\cite{Jang_merging:15}. In order to explore this limiting scenario, we carry out numerical simulations assuming the same phase amplitude $A = 0.3$~rad as our experiments on the 50 element data sequence shown in Fig.~\ref{fig:3}. Specifically we model our system using the normalized Lugiato-Lefever mean-field equation (see, e.g. Ref~\cite{LLE:13}) which render our results generally applicable to any passive Kerr resonator. To this end, we introduce the dimensionless frequency parameter,~$f' = f\sqrt{|\beta_2|L/(2\alpha)}$, where $\alpha$ is the fraction corresponding to half the roundtrip percentage power loss (for our ring cavity $\alpha = 0.146$). These simulations show that phase modulation re-timing remains effective provided the normalised bit-rate remains below $f' = 0.26$. For the cavity parameters presented in this letter this corresponds to a maximum bit rate of 95~Gb/s. At bit rates above this limit adjacent CSs interact strongly compromising the fidelity of the buffer. The other two key parameters of the buffer, storage time and buffer length, possess no intrinsic limits. In practice, storage time depends on the accuracy, and stability, with which the detuning of the driving field, and the mismatch between the modulation frequency and the FSR, can be maintained. Buffer length simply scales with fiber length. This work has already demonstrated that, for a 100~m cavity, storage times in excess of 1~minute are feasible. At the (numerically estimated) maximum bit-rate of 95~Gb/s, such a cavity could support more than 45,000~optical bits.

In conclusion, we have demonstrated an all-optical buffer capable of storing 4736~bits of optical data at a data rate of 10~Gb/s. The system is based on temporal cavity solitons written into a nonlinear fiber ring. The maximum data rate of this buffer was set by the speed of the electronic components used. In theory, data rates approaching 100~Gb/s should be possible before soliton-to-soliton interactions start to overwhelm the re-timing behavior of the phase modulated driving beam. We believe this work clearly demonstrates the exciting potential of temporal cavity solitons for use in all-optical storage and signal-processing systems.

\section*{Funding Information}
Marsden fund of the Royal Society of New Zealand; Rutherford Discovery Fellowships of the Royal Society of New Zealand.

\bigskip


\end{document}